\documentclass[a4paper,twocolumn]{article}
\usepackage{graphicx}
\usepackage{amsmath}

\begin{document}

\title{Quantum capacity of lossy channel with additive classical Gaussian noise : a
perturbation approach}
\author{Xiao-yu Chen \\
{\small {College of Information and Electronic Engineering, Zhejiang
Gongshang University, Hangzhou, 310018, China}}}
\date{}
\maketitle

\begin{abstract}
For a quantum channel of additive Gaussian noise with loss, in the general
case of $n$ copies input, we show that up to first order perturbation, any
non-Gaussian perturbation to the product thermal state input has a less
quantum information transmission rate when the input energy tend to
infinitive.

PACS number(s): 03.67.-a, 42.50.Dv, 89.70.+c
\end{abstract}

\section{Introduction}

Quantum capacity exhibits a kind of nonadditivity \cite{DiVincenzo} that
makes it extremely hard to deal with. The first example with calculable
quantum capacity is quantum erasure channel\cite{Bennett}. Other examples
are dephasing qubit channel\cite{Devetak0}, amplitude damping qubit channel%
\cite{Giovannetti}, and continuous variable lossy channel\cite{Wolf}, where
the channels are either degradable or anti-degradable. Anti-degradable
channel has null quantum capacity due to no clone theorem\cite{Caruso}.
Degradable channel is a channel that when the sender transmits an unknown
quantum state to the receiver with some quantum information leaks to the
environment, the receiver can reconstruct what the environment received from
the state himself received. Degradable quantum channels were first
introduced in Ref. \cite{Devetak0} where it was shown that their quantum
capacity $Q$ can be expressed in terms of the single letter formula of $%
Q=\max \{0,I_c(\sigma ,\mathcal{E})\}.$ Where the coherent information (CI) $%
I_c(\sigma ,\mathcal{E})=S(\mathcal{E}(\sigma ))-S(\sigma ^{QR^{\prime }})$
\cite{Schumacher} \cite{Lloyd}. Here $S(\varrho )=-$Tr$\varrho \log
_2\varrho $ is the von Neumann entropy, $\sigma $ is the input state, the
application of the channel $\mathcal{E}$ results the output state $\mathcal{E%
}(\sigma )$; $\sigma ^{QR^{\prime }}=$ $(\mathcal{E}\otimes \mathbf{I}%
)(\left| \psi \right\rangle \left\langle \psi \right| )$, with $R$ referred
to the 'reference' system\cite{Schumacher} (the system under process is $Q$
system with annihilation and creation operators $a$ and $a^{\dagger }$, we
denote $\sigma ^Q$ as $\sigma $ for simplicity), $\left| \psi \right\rangle $
is the purification of the input state $\sigma $. If a channel is not
degradable, the regulation procedure should be applied to the quantum
capacity, which is \cite{Devetak}\cite{Barnum}\cite{Horodecki}
\begin{equation}
Q=\lim_{n\rightarrow \infty }\sup_{\sigma _n}\frac 1nI_c(\sigma _n,\mathcal{E%
}^{\otimes n}).  \label{wave0}
\end{equation}

Bosonic Gaussian channels \cite{Holevo} include all the physical
transformations which preserve ''Gaussian character'' of the
transmitted signals and can be seen as the quantum counterpart of
the Gaussian channels in the classical information theory. A full
classification of one-mode Bosonic Gaussian channels was presented
in Ref. \cite{Caruso1}, where a very useful channel which is lossy
accompanied by additive classical Gaussian
noise is classified as weak degradable. As stressed by the Authors of Ref.%
\cite{Caruso1}, weak degradability is quite different from
degradability. Thus for such a channel, a single letter formula of
$Q$ may not be available. The regulation formula of (\ref{wave0})
is needed.

\section{The channel and the single letter formula}

The lossy channel with additive classical Gaussian noise can be described by
\begin{eqnarray}
\mathcal{E}(\sigma ) &=&\frac 1{N_n}\int \frac{d^2\alpha }\pi \exp (-\left|
\alpha \right| ^2/N_n)  \nonumber \\
&&\times D\left( \alpha \right) tr_E[U(\sigma \otimes \left| 0\right\rangle
_E\left\langle 0\right| _E)U^{\dagger }]D^{\dagger }(\alpha ),  \label{wave2}
\end{eqnarray}
where $N_n$ specifies the additive classical Gaussian noise, with $D\left(
\alpha \right) =$ $\exp (\alpha a^{\dagger }-\alpha ^{*}a)$ is the
displacement operator. The unitary operator $U=\exp [\theta (aa_E^{\dagger
}-a^{\dagger }a_E)]$ with $a_E$ the annihilation operator of the
environment, and $\eta =\cos ^2\theta $ is the quantum efficiency. Any
quantum state $\sigma $ can be equivalently specified by its characteristic
function $\chi _\sigma (\mu )=tr[\sigma \mathcal{D}(\mu )]$, and inversely $%
\mathcal{\sigma }=\int [\prod_i\frac{d^2\mu _i}\pi ]\chi _{\mathcal{\sigma }%
}(\mu )\mathcal{D}(-\mu )$. The characteristic function of the output state $%
\sigma ^{\prime }=\mathcal{E}(\sigma )$ is \cite{Holevo}
\[
\chi _\sigma ^{\prime }(\mu )=\chi _\sigma (\sqrt{\eta }\mu )e^{-(N_n+\frac{%
1-\eta }2)\left| \mu \right| ^2}.
\]
Thus in the form of characteristic function, the additive property
of the classical Gaussian noise $N_n$ is quite apparently.

A single mode thermal state $\rho $ has a characteristic function of the
form $\chi (\mu )=\exp [-(N+\frac 12)\left| \mu \right| ^2]$, and we have $%
\rho =\int \frac{d^2\mu }\pi \chi (\mu )\mathcal{D}(-\mu )=$ $%
(1-v)v^{a^{\dagger }a}$ with $v=N/(N+1),$[conventionally in the following, $%
v_x=N_x/(N_x+1)$], where $N$ is the average photon number. The noisy lossy
state is $\rho ^{\prime }=\mathcal{E}(\rho )=(1-v^{\prime })v^{\prime
a^{\dagger }a},$ with average photon number $N^{\prime }=\eta N+N_n.$ We
denote the annihilation and creation operators of the 'reference' $R$ system
as $b$ and $b^{\dagger }$. For thermal state input $\rho $, by a proper
symplectic transformation, the joint output state $\rho ^{QR^{\prime }}$ can
be transformed to a direct product of two thermal states with average photon
numbers $N_A$ and $N_B,$ respectively, where $N_{A,B}=\frac 12[D\pm
(N^{\prime }-N)-1]$ with $D=\sqrt{(N^{\prime }+N_n+1)^2-4\eta N(N+1)}$,
yields the coherent \cite{Holevo}
\begin{equation}
I_c(\rho ,\mathcal{E})=g(N^{\prime })-g(N_A)-g(N_B),  \label{wave4}
\end{equation}
where $g(s)=(s+1)\log _2(s+1)-s\log _2s$ is the bosonic entropy function,
and
\begin{equation}
N=N_B\cosh ^2r+(N_A+1)\sinh ^2r,  \label{wave5}
\end{equation}
with $r$ is the parameter of the symplectic transformation, and $\tanh 2r=2%
\sqrt{\eta N(N+1)}/(N^{\prime }+N_n+1).$ Based on the coherent information
of single mode thermal state input, the quantum capacity of the channel has
been conjectured as\cite{Holevo}
\begin{eqnarray}
Q &=&\max \{0,\lim_{N\rightarrow \infty }I_c(\rho ,\mathcal{E})\}  \nonumber
\\
&=&\max \{0,\log _2\left| \eta \right| -\log _2\left| 1-\eta \right|
-g\left| \frac{N_n}{1-\eta }\right| \},  \label{wave1}
\end{eqnarray}
Apart from the regulation, this single letter formula is doubtful
for the input state is quite special. The procedure of
maximization over all continuous variable input state (Gaussian or
non-Gaussian ) has not been taken yet. In the next section, we
will prove that for all single mode Gaussian state input the
single letter quantum capacity is really given by Eq.
(\ref{wave1}).

\section{Gaussian state input to the one-mode channel}

We now consider a single mode Gaussian state $\rho _G$ (which comprises
thermal noise state as its special case) input to the single use of the
channel. A single mode Gaussian state is described by its real correlation
matrix $\alpha $ (we drop the first moments for they can be removed by local
operations) which can be generated from that of thermal noise state $\rho $
with a symplectic transformation \cite{Holevo1}. We have $\alpha =\left[
\begin{array}{ll}
\alpha _{qq} & \alpha _{qp} \\
\alpha _{qp} & \alpha _{pp}
\end{array}
\right] $ with $\det (\alpha )=(N+\frac 12)^2$ . The energy of the Gaussian
state is $E=Tr[(a^{\dagger }a+\frac 12)\rho _G]=\frac 12(\alpha _{qq}+\alpha
_{pp}).$ For a Gaussian state input $\rho _G$, the output $\rho _G^{\prime }$
and the joint output state $\rho _G^{QR^{\prime }}$ are still Gaussian. The
symplectic eigenvalues \cite{Holevo} of these states can be obtained. The
coherent information is
\begin{equation}
I_c(\rho _G,\mathcal{E})=g(d_0-\frac 12)-g(d_1-\frac 12)-g(d_2-\frac 12),
\end{equation}
with
\begin{eqnarray}
d_0 &=&\sqrt{N_n^{\prime 2}+2\eta EN_n^{\prime }+\eta ^2E^2x},\text{ } \\
d_{1,2} &=&\sqrt{\frac 12(X\pm \sqrt{X^2-4Y})},
\end{eqnarray}
where $N_n^{\prime }=N_n+\frac{1-\eta }2,$ $X=N_n^{^{\prime }2}+2\eta
EN_n^{\prime }+\frac \eta 2+(1-\eta )^2E^2x,$ $Y=\frac 12\eta EN_n^{\prime
}+\frac 1{16}\eta ^2+E^2N_n^{\prime 2}x,$ with $x=y^{-2}$, and $y=E/(N+\frac
12)$. With the condition $\det (\alpha )=(N+\frac 12)^2,$ it is not
difficult to prove that $y$ has its global minimum $y=1$ when $\alpha
_{qq}=\alpha _{pp}=N+\frac 12,$ $\alpha _{qp}=0$. The derivative of the
coherent information with respect to $x$ is $\frac{dI_c(\rho _G,\mathcal{E})%
}{dx}=f(d_0)-f(d_1)-f(d_2),$ with $f(z)=\frac{dg(z-1/2)}{dx}=\frac 1{2z}\log
_2\frac{z+1/2}{z-1/2}\frac{d(z^2)}{dx}.$ When $E\rightarrow \infty ,$ we
have $d_0\symbol{126}\eta E\sqrt{x}\rightarrow \infty ,$ $d_1\symbol{126}%
(1-\eta )E\sqrt{x}\rightarrow \infty ,$then
\begin{equation}
f(d_0)-f(d_1)=-\frac{N_n^{\prime }}{xE\ln 2}[\frac 1\eta -\frac \eta
{(1-\eta )^2}].
\end{equation}
Note that even when $E\rightarrow \infty ,$ we have $d_2=N_n^{\prime
}/(1-\eta ),$ thus
\begin{equation}
\left. f(d_2)\right| _{E\rightarrow \infty }=\frac{3\eta [N_n^{\prime
2}-\frac 14(1-\eta )^2]}{4x^2(1-\eta )^3}\log _2\frac{N_n^{\prime }+\frac
12(1-\eta )}{N_n^{\prime }-\frac 12(1-\eta )},
\end{equation}
which is always positive for nonzero noise $N_n$. Hence for sufficiently
large input energy $E,$ we have $\frac{dI_c(\rho _G,\mathcal{E})}{dx}<0.$
While $x$ has its global minimum value $x=1,$ so the coherent information
achieves its maximum at $x=1$ which corresponds to thermal noise state
input. Hence we can conclude that for sufficient large but definite input
energy, the one-shot quantum information capacity of the channel is achieved
by thermal noise state input of all Gaussian state inputs.

\section{Perturbation to the $n$ use of the channel}

In the $n$ use of the channel with an input Gaussian state $\rho _n,$the
algebraic equations of the symplectic eigenvalues \cite{Holevo} are not
analytically solvable. And for non-Gaussian state input, it is even worse in
calculating the coherent information. So, in this paper, we turn to
perturbation of the conjectured extremal state of the product thermal state $%
\rho ^{\otimes n}.$ To treat the problem with perturbation theory, we need
the following lemmas:

\textit{Lemma : }(1)\textit{\ }$(\mathcal{E}\otimes \mathbf{I)}(a^{\dagger
k}\rho ^{QR}a^m)\mathbf{=}v^{-(k+m)/2}b^k\rho ^{QR^{\prime }}b^{\dagger m}$;
(2) $(\mathcal{E}\otimes \mathbf{I)}(a^k\rho ^{QR}a^{\dagger m})\mathbf{=}%
v^{(k+m)/2}b^{\dagger k}\rho ^{QR^{\prime }}b^m.$

\textit{proof: } (1) Both of the characteristic functions of the lhs and the
rhs are equal to $(1-v)\exp [-\frac 12\left| \mu _b\right| ^2+(\eta -\frac
12-N_n)\left| \mu _a\right| ^2]\cdot $ $\int \frac{d^2\alpha }\pi \alpha
^{*k}\alpha ^m\exp [-(1-v)\left| \alpha \right| ^2$ $+(\sqrt{v}\mu _b-\sqrt{%
\eta }\mu _a^{*})\alpha $ $-(\sqrt{v}\mu _b^{*}-\sqrt{\eta }\mu _a)\alpha
^{*}]$. (2) can be proved similarly.

We consider the first order multi-mode perturbation to the input product
thermal state $\rho ^{\otimes n}$. A typical case is $\chi _{n\varepsilon }(%
\mathbf{\mu })=\chi _n(\mathbf{\mu })[1+\varepsilon (c\mu _1^{k_1}\mu
_2^{k_2}\cdots \mu _n^{k_n}\mu _1^{*l_1}\mu _2^{*l_2}\cdots \mu _n^{*l_n}$ $%
+c^{*}\mu _1^{*k_1}\mu _2^{*k_2}\cdots \mu _n^{*k_n}\mu _1^{l_1}\mu
_2^{l_2}\cdots \mu _n^{l_n})]$ with $\sum_{i=1}^nk_i=\sum_{i=1}^nl_i=m$ (the
requirement of first order perturbation). We may denote the perturbation as $%
(\mathbf{k},\mathbf{l}),$ with vectors $\mathbf{k}=(k_1,k_2,\cdots ,k_n),$ $%
\mathbf{l}=(l_1,l_2,\cdots ,l_n).$ The perturbed input state is $\rho
_{n\varepsilon }=\rho ^{\otimes n}+\phi .$ The general form of the
perturbation should be a linear combination of this typical $\phi .$ We will
prove that each $\phi $ contributes independently to the coherent
information a negative quantity. To simplify the calculation, we introduce a
generation function $I_\phi \left( \mathbf{\tau ,\sigma }\right) =\int
\left[ \Pi _i\frac{d^2\mu _i}\pi \right] \chi _n(\mathbf{\mu })D(-\mathbf{%
\mu })\exp [\mathbf{\mu \cdot \tau +\mu }^{*}\cdot \mathbf{\sigma }],$ then
\begin{equation}
\phi =\left( c\frac{\partial ^{2m}I_\phi \left( \mathbf{\tau ,\sigma }%
\right) }{\Pi _i(\partial \tau _i^{k_i}\partial \sigma _i^{l_i})}+c^{*}\frac{%
\partial ^{2m}I_\phi \left( \mathbf{\tau ,\sigma }\right) }{\Pi _i(\partial
\tau _i^{l_i}\partial \sigma _i^{k_i})}\right) _{\mathbf{\tau =\sigma =0}}.
\end{equation}
It has been proved in Ref. \cite{Chen} that the perturbation to the entropy
is
\begin{equation}
S(\rho _{n\varepsilon })-S(\rho ^{\otimes n})=-\frac 12\varepsilon ^2Tr(\phi
^2/\rho ^{\otimes n})+o(\varepsilon ^3).
\end{equation}
For
\begin{equation}
Tr[I_\phi \left( \mathbf{\tau ,\sigma }\right) I_\phi \left( \mathbf{\tau }%
^{\prime }\mathbf{,\sigma }^{\prime }\right) /\rho ^{\otimes n}]=\exp [-%
\frac{\mathbf{\tau \cdot \sigma }^{\prime }}N-\frac{\mathbf{\tau }^{\prime }%
\mathbf{\cdot \sigma }}{N+1}],  \label{wave13}
\end{equation}
Thus
\begin{equation}
Tr(\phi ^2/\rho ^{\otimes n})=2c_0\frac{\Pi _i(k_i!l_i!)}{[N(N+1)]^m}.
\end{equation}
where $c_0=\left| c\right| ^2$ for $\mathbf{k\neq l}$ and $c_0=4c_R^2$ for $%
\mathbf{k=l,}$ $c_R$ is the real part of $c.$ The perturbation to the
entropy of the output state $\rho ^{\prime \otimes n}$ is
\begin{equation}
S(\rho _{n\varepsilon }^{\prime })-S(\rho ^{\prime \otimes n})=-\varepsilon
^2c_0\frac{\Pi _i(k_i!l_i!)}{[N^{\prime }(N^{\prime }+1)]^m}+o(\varepsilon
^3).
\end{equation}
Eq.(\ref{wave13}) exhibits that any intercross item of $Tr(\phi \phi
^{\prime }/\rho ^{\otimes n})$ type will be nullified for $(\mathbf{k},%
\mathbf{l})\neq $ $(\mathbf{k}^{\prime },\mathbf{l}^{\prime })$. Thus each
perturbation item contributes to the entropy separately.

The perturbation to the joint $QR$ state is more sophisticated. We may
express the perturbed joint input state as $\rho _{n\varepsilon }^{QR}=\rho
^{QR\otimes n}+\varepsilon \Phi ,$ and $\Phi =\frac 12(\Phi _0+\Phi
_0^{\dagger }).$ The generation function of $\Phi _0$ is \cite{Chen}
\begin{equation}
I_{\Phi _0}=\exp \left[ \frac{\mathbf{\sigma \cdot (\tau -a}^{\dagger }%
\mathbf{)}}{N+1}\right] \exp \left[ \frac{\mathbf{\tau \cdot a}}N\right]
\rho ^{QR\otimes n}.
\end{equation}
The action of the channel then is
\begin{equation}
I_{\Phi _0}^{\prime }=(\mathcal{E}\otimes \mathbf{I)}I_{\Phi _0}=\exp (p%
\mathbf{\tau \cdot b}^{\dagger })\exp [\frac{\mathbf{\tau \cdot \sigma }}{N+1%
}-p\mathbf{\sigma \cdot b]}\rho ^{QR^{\prime }\otimes n}
\end{equation}
according to the lemma, where $p=[N(N+1)]^{-1/2}$, and we have used the fact
that $\exp (\frac{\tau a}N)\rho ^{QR}=$ $\exp (p\tau b^{\dagger })\rho ^{QR}$%
. The contribution to the entropy should be evaluated in the eigenbasis of $%
\rho ^{QR^{\prime }\otimes n}$. We may denote the subspace of $\rho
^{QR^{\prime }\otimes n}$as$\left| i,\mathbf{i},j,\mathbf{j}\right\rangle
^{\prime }$ which has eigenvalue $\lambda _{ij}=(1-v_A)^n(1-v_B)^nv_A^iv_B^j$
$,$ where $\mathbf{i}=(i_1,i_2,\ldots ,i_n),$ $i=\sum_{s=1}^ni_s,\mathbf{j}%
=(j_1,j_2,\ldots ,j_n),$ $j=\sum_{s=1}^nj_s.$ In this subspace, we denote $%
\Phi _0^{\prime }$ as $M_{ij},$ the elements of $M_{ij}$ are $\left\langle i,%
\mathbf{i},j,\mathbf{j}\right| ^{\prime }\Phi _0^{\prime }\left| i,\mathbf{i}%
^{\prime },j,\mathbf{j}^{\prime }\right\rangle ^{\prime },$ the sum of the
square of the eigenvalue of is $TrM_{ij}^2$. We obtain the contribution to
the entropy by first evaluating $\left\langle i,\mathbf{i;}j,\mathbf{j}%
\right| ^{\prime }I_{\Phi _0}^{\prime }\left| i,\mathbf{i}^{\prime }\mathbf{,%
}j,\mathbf{j}^{\prime }\right\rangle ^{\prime }=\left\langle i,\mathbf{i;}j,%
\mathbf{j}\right| V^{\dagger \otimes n}I_{\Phi _0}^{\prime }V^{\otimes
n}\left| i,\mathbf{i}^{\prime }\mathbf{,}j,\mathbf{j}^{\prime }\right\rangle
,$ which is $\exp [\mathbf{\tau \cdot \sigma /(}N+1)]$ $\left\langle i,%
\mathbf{i;}j,\mathbf{j}\right| \exp [p\mathbf{\tau \cdot (b}^{\dagger }\cosh
r+\mathbf{a}\sinh r)]$ $\exp [-p\mathbf{\sigma \cdot (b}\cosh r+\mathbf{a}%
^{\dagger }\sinh r)\left| i,\mathbf{i}^{\prime };j,\mathbf{j}^{\prime
}\right\rangle .$ Here $V$ is the unitary transformation that transforms $%
\rho ^{QR^{\prime }}$ to its product form of $A$ and $B$ parts. We expand
the exponent of the operators to drop the terms that do not keep the total
particle numbers in $A$ and $B$ parts respectively. Denote
\begin{eqnarray}
I_B(\mathbf{\tau ,\sigma }) &=&\sum_{k=0}^\infty \frac{(-1)^k}{k!^2}(p\cosh
r)^{2k}(\mathbf{\tau \cdot b}^{\dagger })^k(\mathbf{\sigma \cdot b})^k, \\
I_A(\mathbf{\tau ,\sigma }) &=&\sum_{k=0}^\infty \frac{(-1)^k}{k!^2}(p\sinh
r)^{2k}(\mathbf{\tau \cdot a})^k(\mathbf{\sigma \cdot a}^{\dagger })^k.
\end{eqnarray}
Then in the calculation of the contribution to the entropy become a trace on
the whole space, the restriction on the subspace is removed. We have the
generation function
\begin{eqnarray}
F &=&\exp [(\mathbf{\tau \cdot \sigma +\tau }^{\prime }\mathbf{\cdot \sigma }%
^{\prime }\mathbf{)/(}N+1)]  \nonumber \\
&&\times Tr[(I_B(\mathbf{\tau ,\sigma })I_B(\mathbf{\tau }^{\prime }\mathbf{%
,\sigma }^{\prime })/\rho _B^{\otimes n}]  \nonumber \\
&&\times Tr[I_A(\mathbf{\tau ,\sigma })I_A(\mathbf{\tau }^{\prime }\mathbf{%
,\sigma }^{\prime })/\rho _A^{\otimes n}]  \nonumber \\
&=&\sum_{m=0}^\infty \sum_{l=0}^m\frac{B^jA^{m-j}}{[j!(m-j)!]^2[N(N+1)]^{2m}}
\nonumber \\
&&\times (\mathbf{\tau \cdot \sigma }^{\prime })^m(\mathbf{\tau }^{\prime }%
\mathbf{\cdot \sigma })^m.  \label{wave14}
\end{eqnarray}
with $B=N_B(N_B+1)\cosh ^4r,$ $A=N_A(N_A+1)\sinh ^4r$. Where we have used
another generation function in evaluating $F,$ and at the final step we
exchange the orders of summation and make use of Eq.(\ref{wave5}) to conceal
the factor $\exp [(\mathbf{\tau \cdot \sigma +\tau }^{\prime }\mathbf{\cdot
\sigma }^{\prime }\mathbf{)/(}N+1)].$ The generation function is about that
the two ingredient both come from $\Phi _0,$ if both come from $\Phi
_0^{\dagger },$the result will be the same. In the case intercross of $\Phi
_0$ and $\Phi _0^{\dagger },$ we should substitute $(\mathbf{\tau \cdot
\sigma }^{\prime })^m(\mathbf{\tau }^{\prime }\mathbf{\cdot \sigma })^m$ in
Eq.(\ref{wave14}) by $(\mathbf{\tau \cdot \tau }^{\prime })^m(\mathbf{\sigma
}^{\prime }\mathbf{\cdot \sigma })^m$.

\begin{eqnarray}
\sum_{ij}\frac{Tr[\frac 12(M_{ij}+M_{ij}^{\dagger })]^2}{\lambda _{ij}}
&=&2c_0\frac{\Pi _i(k_i!l_i!)}{[N(N+1)]^{2m}}  \nonumber \\
&&\times \sum_{l=0}^m\binom mj^2B^jA^{m-j}.
\end{eqnarray}
When $N^{\prime }<N,$ that is $N>N_n/(1-\eta ),$ we have $\frac{N(N+1)}{%
N^{\prime }(N^{\prime }+1)}>1$. With Eq.(\ref{wave5}), it is not difficult
to prove that
\[
\frac 1{N^m(N+1)^m}\sum_{l=0}^m\binom mj^2B^jA^{m-j}<1.
\]
Thus
\[
\lim_{N\rightarrow \infty }[I_c(\rho _{n\varepsilon },\mathcal{E}^{\otimes
n})-I_c(\rho ^{\otimes n},\mathcal{E}^{\otimes n})<0.
\]
Eq.(\ref{wave14}) indicates that each perturbation term contributes to the
entropy separately. In expanding $\rho _{n\varepsilon }^{QR}$, there is the $%
\varepsilon ^2$ term which will also contributes to the entropy up to $%
\varepsilon ^2$. But this is negligible at large input energy.

\section{Conclusions}

We have shown that all first order perturbation to the input
product identical thermal state can only decrease the coherent
information at large input energy in the most general case of $n$
use of channel. Any perturbation to the input state contributes
the coherent information independently. A linear combination of
perturbations to the input state results a linear combination of
the perturbations of the coherent information. In the sense of
first order perturbation, the channel capacity of additive
Gaussian quantum channel with loss is described by the long
standing conjecture formula (\ref{wave1}).

\textit{\ }Funding by the National Natural Science Foundation of China
(Grant No. 10575092), Zhejiang Province Natural Science Foundation (Grant
No. RC104265) are gratefully acknowledged.


\begin{thebibliography}{99}
\bibitem{DiVincenzo}  D. P. DiVincenzo, P. W. Shor and J. A. Smolin,
Phys.Rev. A \textbf{57}, 830 (1998).

\bibitem{Bennett}  C. H. Bennett, D. P. Divincenzo and J. A. Smolin, Phys.
Rev. Lett. \textbf{78}, 3217 (1997).

\bibitem{Devetak0}  I. Devetak, and P. W. Shor, Comm. Math. Phys. \textbf{256%
}, 287(2005); arXiv:quant-ph/0311131.

\bibitem{Giovannetti}  V. Giovannetti, R. Fazio, Phys. Rev. A, \textbf{71},
032314 (2005).

\bibitem{Wolf}  M. M. Wolf, D. P\textexclamdown \"{a}erez-Garc%
\textexclamdown \"{a}ia, G. Giedke, Phys. Rev. Lett. \textbf{98},
130501(2007).

\bibitem{Caruso}  F. Caruso, V. Giovannetti, Phys. Rev. A \textbf{74},
062307 (2006)

\bibitem{Schumacher}  B. Schumacher, Phys. Rev. A \textbf{54,} 2614 (1996);
B. Schumacher and M. A. Nielsen, Phys. Rev. A \textbf{54,} 2629 (1996).

\bibitem{Lloyd}  S. Lloyd, Phys. Rev. A \textbf{55, }1613 (1997).

\bibitem{Devetak}  I. Devetak, IEEE Trans. inf. Theory \textbf{51},
44(2005); I. Devetak and A. Winter, Proc. R. Soc. Lond. A, \textbf{461}, 207
(2005).

\bibitem{Barnum}  H. Barnum, M. Knill and M. A. Nielsen, IEEE Trans. Inf.
Theory, \textbf{46, }1317 (2000).

\bibitem{Horodecki}  M. Horodecki, P. Horodecki and R. Horodecki, Phys. Rev.
Lett. \textbf{85}, 433 (2000).

\bibitem{Holevo}  A. S. Holevo and R. F. Werner, Phys.Rev. A \textbf{63},
032312 (2001).

\bibitem{Holevo1}  A. S. Holevo, M. Sohma and O. Hirota, Phys.Rev. A,
\textbf{59}, 1820 (1999).

\bibitem{Caruso1}  F. Caruso, V. Giovannetti, A. S. Holevo, New J. Phys.
\textbf{8, }310 (2006).

\bibitem{Chen}  X. Y. Chen, arXiv: 0709.2047v1
\end{thebibliography}
\end{document}